\documentclass[aps,pra,twocolumn,superscriptaddress]{revtex4-2}
\usepackage{bm}
\usepackage{graphicx}
\usepackage{amssymb,amsmath,amsbsy,amsgen,amsfonts}
\usepackage{dcolumn}
\usepackage{amsthm}
\usepackage{mathrsfs}
\usepackage{latexsym}
\usepackage{array}
\usepackage{amstext}
\usepackage{epsfig}
\usepackage{epstopdf} 

\usepackage{color}
\usepackage{units}
\usepackage{calrsfs}

\newcommand{\be}{\begin{equation}}
\newcommand{\ee}{\end{equation}}
\newcommand{\ba}{\begin{array}}
\newcommand{\ea}{\end{array}}
\newcommand{\bqa}{\begin{eqnarray}}
\newcommand{\eqa}{\end{eqnarray}}

\begin{document}

\title{Enhancement of Quantum Excitation Transport by Photonic Nonreciprocity}

\author{S. Ali Hassani Gangaraj} \email{ali.gangaraj@gmail.com}
\address{Department of Electrical and Computer Engineering, University of Wisconsin-Madison, Madison, WI 53706, USA}

\author{Lei Ying} 
\address{Department of Physics, Interdisciplinary Center for Quantum Information, State Key Laboratory of Modern Optical Instrumentation, and Zhejiang Province Key Laboratory of Quantum Technology and Device, Zhejiang University, Hangzhou 310027, China}

\author{Francesco Monticone} 
\address{School of Electrical and Computer Engineering, Cornell University, Ithaca, NY 14853, USA}

\author{Zongfu Yu} 
\address{Department of Electrical and Computer Engineering, University of Wisconsin-Madison, Madison, WI 53706, USA}

\date{\today}

\begin{abstract}

Enhanced interaction between two two-level emitters (e.g., atoms) by nonreciprocal photonic media can be of benefit to broad areas, from quantum information science to biological detection. Here we provide a detailed analysis on why nonreciprocal photon-mediated interaction enhances inter-atomic excitation transport efficiency. We investigate a system consisting of two two-level emitters embedded in a generic photonic environment. By comparing symmetric and asymmetric photon-exchange, we analytically show that breaking electromagnetic reciprocity makes it possible for the cooperative decay rate to exceed the spontaneous decay rate even in a translation-invariant homogeneous system. This means that the excitation of an emitter must decay mostly into the other emitter rather than leaking and dissipating into the reservoir photonic modes. We also provide an example where a chain of two-level emitters dominantly interact via the reciprocal modes of a plasmonic waveguide. We then show that breaking reciprocity in such a system via driving a DC current through the plasmonic material can drastically increase the probability of photon emission from one emitter to another, leading to an order-of-magnitude enhancement in quantum energy-transport efficiency.

\end{abstract}

\maketitle


\section{Introduction}

Maintaining entanglement and quantum superposition between two separated two-level (TL) ``atoms'' in the face of interactions with the surrounding reservoir is one of the fundamental aspects of quantum technologies - in computing, sensing, quantum internet and photon transduction - to outperform their classical counterparts \cite{Q_decoherence}. In this regard, one of the main approaches to address this challenge is strengthening the quantum emitters photon-exchange process (also known as inter-atomic energy transport) to beat photon leakage into the environment. In addition, the interest in efficient micro-scale energy transport plays a key role in a wide range of biological and non-biological systems, from photosynthesis in plants and bacteria \cite{photosynthesis_1,photosynthesis_2} to many other applications in lighting \cite{lighting_1, lighting_2,lighting_3}, photovoltaics \cite{Photovoltaic_1,Photovoltaic_2}, F\"orster energy transfer \cite{Foster_1,Foster_2} and sensing \cite{sensing}.

For this purpose, significant effort has been devoted to understand the mechanism of energy transport and explore how to enhance it within open quantum systems \cite{OQS_1,OQS_2,OQS_3,OQS_4,OQS_5,OQS_6,OQS_7,OQS_8,Pupillo,Poddubny,citekey,Roque,Ying2019,Ying2022}. As a prominent example, plasmonic waveguide platforms can be mentioned where inter-atomic photon exchange is mediated by surface plasmon polaritons (SPPs) \cite{Sorensen_2,1DSPP,Harvesting,Moreno,Barnes,OpticsExpress_Hassani,Hassani_PRL_EP,Hassani_PRB}. These systems provide strong radiative coupling due to tight confinement of fields. In this context a successful strategy for enhancing the interaction is imposing asymmetry in photon (quantum of excitation) exchange \cite{Zoller_chiral,Zoller_spin,Zoler_cascade,Zoller_dimer,chiral_route,Sorensen,Iorsh,Makhonin}, and taking advantage of this asymmetric energy transport.

Asymmetric photon exchange can be realized via either chiral or nonreciprocal interfaces. In chiral interfaces, the coupling between photon and TL emitter depends on the photon propagation direction and the polarization of the emitter dipole moment. Consequently, although the photonic waveguide itself may respect Lorentz reciprocity, chirality breaks the symmetry of photon emission from the TL emitter into right and left propagating waveguide modes \cite{Zoller_chiral}. Chiral interaction between a pair of emitters provides unique quantum many-body systems, for instance the steady state of the system can be a pure entangled many-atom state in the case of chiral coupling to a reciprocal waveguide \cite{Zoler_cascade,Zoller_chiral,Zoller_dimer,Zoller_spin}. On the other hand, in nonreciprocal interfaces, symmetry is broken at a more fundamental level, leading to inherent asymmetry between counter-propagating modes in a photonic waveguide. 

Electromagnetic nonreciprocity has been a functional tool for applications in classical wave physics ranging from microwave isolators and circulators to unidirectional optical signal routing. With the advent of quantum technologies, nonreciprocal photon transduction is also becoming beneficial for quantum communications. For example reference \cite{PRA-Hassani} studied pumped entanglement of TL emitters in the vicinity of photonic topological insulators. The resulting entanglement has been shown to be robust to physical defects due to unidirectional and defect-immune SPPs supported at their interface. In another study \cite{PRL-Hassani}, it has been shown that coupling a chain of TL emitters to the nonreciprocal modes of PTIs enhances the energy-transport efficiency compared to reciprocal plasmonic waveguides. In Ref.~\cite{PRL-Hassani}, it has been argued that such a large enhancement is due to large dissipative decay rate and peculiar properties of topological edge states.

Here, building upon these works, we provide a simple yet comprehensible discussion on why imposing asymmetry - by any means - on photon exchange process can increase the probability of photon emission from one atom to another, which eventually enhances the efficiency of inter-atomic quantum excitation transport. The idea driving the present work is exploring the advantageous properties of systems with broken electromagnetic reciprocity for inter-atomic energy transport. Our approach is based on three-dimensional (3D) Green's function and we focus on systems where identical atoms interact with each other ($\Gamma_{11}$ = $\Gamma_{22}$). 

This paper is oranaized as follow: In section II, using 3D Green's function, we discuss the coupling of emitters to the environment modes and their effects upon each other. We elucidate the dynamical evolution of two TL emitters interacting with each other in a generic environment under reciprocity and nonreciprocity assumption. By calculating the probability of finding the atoms in their excited state, we show that violating the Lorentz reciprocity makes this possible that the dissipative decay rate, $ \Gamma_{\mathrm{i} \neq \mathrm{j} } $, exceeds the spontaneous decay rate, $ \Gamma_{\mathrm{i} = \mathrm{j} } $, even in a translation-invariant homogeneous system. This condition is essential to increase the probability of photon emission from one emitter to another. This means the excitation of an emitter decays mostly into the other emitter rather than leaking into the reservoir photonic modes. We discuss that although this is a key ingredient for strong photon exchange, it cannot be effectively achieved unless we break the system reciprocity. We also show that how this condition is linked to the nonzero spatial derivative of the system Green function at the emitter location, and how it improves the fundamental limit in photon transport process. 

In section III we provide an example where two identical atoms interact via a reciprocal 3D plasmonic platform, i.e., an interface between indium antimonide semiconductor (InSb) and vacuum. We investigate how the quantum emitters interact under reciprocity assumption, but then we break the reciprocity by biasing the plasmonic materials with a direct electric current. As explained in \cite{Bliokh,Drifting_electrons}, the drifting electrons produce a Doppler shift in the isotropic material permittivity, i.e., $ \omega \rightarrow \omega - \textbf{k} \cdot \textbf{v}_d $, where $ \textbf{k} $ is the wavevector and $ \textbf{v}_d $ is the electron drift velocity. This linear dependence on the wavevector implies that the material response becomes nonreciprocal. This method of breaking reciprocity is magnet-free, in addition it is compatible with quantum circuits for on-chip integration, detailed discussion are provided in the section III. In this situation the TL emitters interact via nonreciprocal SPPs. We discuss that by breaking reciprocity significant values of photon transport efficiency can be achieved compared to reciprocal interaction. In addition we discuss the possibility of tuning energy transport through an easily accessible parameter, that is the direction of the electron drift velocity.


\section{Competition Between Spontaneous Emission and Dipole-Dipole Interaction}

\subsection{  Collective decay rate and coherent coupling  }

At first,  let us recall the quantum state of a single atom. A single isolated TL atom is stationary and can be described by time-independent Schr\"odinger equation. This means if the atom contains excitation (electron in higher energy level), it does not relax into the ground state \cite{Pelton}. The atom stays excited unless it interacts with the electromagnetic modes of its reservoir and creates a real photon. This process is called spontaneous emission. It is well known that the spontaneous emission rate of a single atom is controlled by its photonic environment \cite{SE-1,SE-2,SE-3,SE-4} and it is associate with the local density of states (LDOS) that counts the number of electromagnetic modes available for the photon emission \cite{SE-2,SE-3,Novotny}. The spontaneous emission of a TL atom is given by the imaginary part of the dyadic Green's function $\mathrm{Im} \mathbb{G}(\mathbf{r}_0, \mathbf{r}_0, \omega_{0}) $ and dipole moment $\boldsymbol{\mu}$ at the location of atom itself. It is given by
\begin{equation}
	\gamma = \frac{2 \omega_{0}^2}{ \hbar \epsilon_0 c^2 } {\boldsymbol{\mu}} \cdot \mathrm{Im} \mathbb{G}(\mathbf{r}_0, \mathbf{r}_0, \omega_{0})  \cdot {\boldsymbol{\mu}},
\end{equation}
where $\omega_0$ is the transition frequency of the TL atom, $\epsilon_0$ is the vacuum permittivity, and $c$ is the speed of light. If the atom is initially excited, the spontaneous emission is the rate at which it emits a photon and decays to its ground state. Therefore, the spontaneous emission is a process that releases the excitation of an atom into the quantized radiation field. The released quantum excitation may eventually vanish due to material absorption. If there is only one atom in the system, then the probability of finding the photon in atom decays as $ P_\mathrm{e} = e^{-\gamma t} $.

In contrast to spontaneous emission, the collective dipole-dipole interaction is differently affected by the reservoir Green's function, namely, it depends on the total Green's function between two points (known as atomic locations) \cite{Welsch,Wubs_Foster}. This interaction occurs between two separated atoms that share common photonic modes \cite{Zubin_1}. Collective dipole-dipole interaction is responsible to transport one quantum of excitation (photon) from one atom (source) to another atom (destination). This process establishes coherent and dissipative couplings between separated atoms and ultimately mediates entanglement.

Now, the interaction between two identical atoms can be described by the resonant dipole-dipole interaction, which can be calculated from the transition matrix element (see Appendix \ref{AppA} for detailed information). Considering the initial and final states as $ \left | \mathrm{i} \right > = \left | e_1,g_2; 0 \right > $ and $ \left | \mathrm{f} \right > = \left | g_1, e_2; 0 \right > $ i.e., transferring one quantum of excitation from the first atom to the second atom while the photon number in the photonic environment is  $ 0 $, the potential strength of the resonant dipole-dipole interaction is given by (see Appendix \ref{AppA}):
\begin{equation}\label{DDI}
\begin{split}
	\frac{M_\mathrm{fi} }{\hbar}
    =-\frac{\omega_{0}^2}{\epsilon_0 c^2} \boldsymbol{\mu}_1 \cdot \mathbb{G} (\mathbf{r}_1, \mathbf{r}_2, \omega_0)  \cdot \boldsymbol{\mu}_2
    =  i\frac{\Gamma_{12}}{2} + g_{12}    ,
\end{split}
\end{equation}
where $ \boldsymbol{\mu}_{1,2}$ are dipole moments of atoms ``$1$'' and ``$2$''. Although the spontaneous emission depends on the Green's function imaginary part at the location of atom itself, the dipole-dipole interaction potential strength obeys the total Green's function propagator. This indicates that the quantum excitation transport depends on the delocalization of the electromagnetic field generated by one quantum emitter at the position of the other one. In other words, while the spontaneous emission decays the excitation into the reservoir quantized field radiation, the dipole-dipole interaction aims to deliver the excitation to the other atom. Therefore, the spontaneous emission is a process that competes with the quantum energy transport. With that being said, it can be concluded that the appearance of the Green’s function in the spontaneous emission and dipole-dipole interaction reveals that although the quantumness of the system is encoded in either the quantum nature of the emitter or the correlation of bosonic field operators, the field propagation is governed by the wave equation. This means the dynamics of the photon exchange process is determined by the Green's function propagator $ \mathbb{G}(\textbf{r}_i, \textbf{r}_j, \omega_{0}) $. Therefore modifying the properties of Green's function provides an effective tool to control and enhance photon transport process.

\subsection{Atomic state dynamics with nonreciprocal dipole-dipole interactions}

To understand how nonreciprocity enhances the photon transport process from the dynamics of TL atoms, we study the excitation transport from one atom to another in a generic photonic environment. In what follows we calculate the probability of finding each atom in the excited state by solving the time evolution of the density matrix under the Born-Markov approximation. The Born-Markov approximation comes from the assumption that the reservoir relaxation time is much faster than the relaxation time of the emitters system, and so the memory effect can be ignored \cite{PRA-Hassani}. This implies that the Green’s function is characterized by a broad spectrum, which can be considered to be flat over the atomic linewidth \cite{Kimble}. Within the rotating wave approximation (the frame rotating with the probe field frequency), the dynamics of the density matrix associated with a chain of $N$ atoms is described by the following expression valid for both reciprocal and nonreciprocal environments in general \cite{PRL-Hassani}:

\begin{align}\label{ME}
	\frac{\partial \rho }{\partial t} & = \frac{-i}{\hbar} \left[ H_\mathrm{sys}, \rho(t)  \right] \nonumber \\& + \sum_{i =1}^{N} \frac{ \Gamma_{ii} }{ 2} \left(  2\sigma_i \rho(t) \sigma_i^{\dagger} -  \sigma_i^{\dagger} \sigma_i \rho(t) -  \rho(t) \sigma_i^{\dagger} \sigma_i   \right) \nonumber \\&
	+ \sum_{i,j}^{i \neq j} \frac{ \Gamma_{ij} }{2 } \left( \left[  \sigma_j \rho(t), \sigma_i^{\dagger} \right] + \left[  \sigma_i, \rho(t)\sigma_j^{\dagger} \right]   \right) \nonumber \\& + \sum_{i,j}^{i \neq j} g_{ij} \left( \left[  \sigma_j \rho(t), -i \sigma_i^{\dagger} \right] + \left[  i \sigma_i, \rho(t)\sigma_j^{\dagger} \right]   \right) \nonumber \\&
	+ \frac{\Gamma_\mathrm{in}}{2} \left(  2\sigma_1^{\dagger} \rho(t) \sigma_1 -  \sigma_1 \sigma_1^{\dagger} \rho - \rho \sigma_1 \sigma_1^{\dagger} \right) \nonumber \\& + \frac{\Gamma_\mathrm{out}}{2} \left(  2\sigma_N \rho(t) \sigma_N^{\dagger} -  \sigma_N^{\dagger} \sigma_N \rho - \rho \sigma_N^{\dagger} \sigma_N \right),
\end{align}
where
\begin{align}
	& \Gamma_{ij} = \frac{2\omega_0^2}{\epsilon_0 \hbar c^2}   \boldsymbol{\mu}_{i} \cdot \mathrm{Im} \left[ \mathbb{G} (\textbf{r}_i, \textbf{r}_j, \omega_{0}) \right] \cdot \boldsymbol{\mu}_{ j} \nonumber \\&
	g_{ij} = \frac{\omega_0^2}{\epsilon_0 \hbar c^2}  \boldsymbol{\mu}_{ i} \cdot \mathrm{Re} \left[ \mathbb{G} (\textbf{r}_i, \textbf{r}_j, \omega_{0}) \right] \cdot \boldsymbol{\mu}_{ j}
\end{align}
are the collective decay rate and the coherent coupling, respectively. When $ i \neq j $, these coupling terms are related to the dipole-dipole potential strength as $ \Gamma_{ij} = -2 \mathrm{Im}\left(M_\mathrm{fi}\right) /\hbar$ and $ g_{ij} = - \mathrm{Re}\left(M_\mathrm{fi}\right) /\hbar$. The parameters $ \Gamma_\mathrm{in},~\Gamma_\mathrm{out} $ are the rates that energy is pumped into or extracted form the first or last atom and $\sigma_i/ \sigma_i^{\dagger}$ are the atomic energy lowering/raising operators. In the scenario without pumping/extraction, we have  $\Gamma_\mathrm{in} = \Gamma_\mathrm{out} = 0$.

\begin{figure*}[bth!]
	\begin{center}
		\noindent
		\includegraphics[width=1.5\columnwidth]{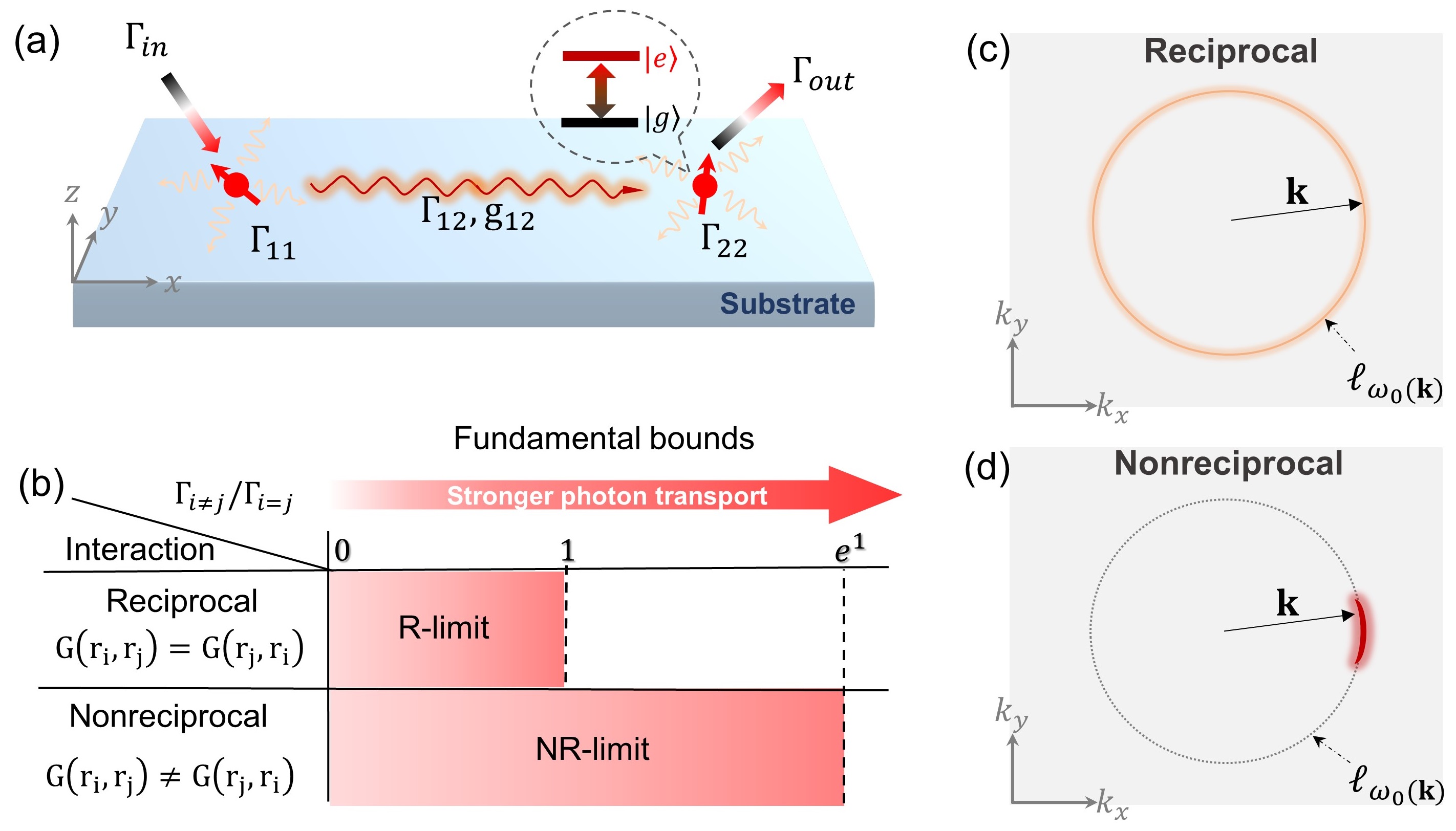}
		\caption{ (a) The system under consideration: Two TL emitters located above a translationally invariant structure. Energy is pumped into the first atom with rate $ \Gamma_\mathrm{in} $ and extracted from the second atom with rate $ \Gamma_\mathrm{out} $. The spontaneous emissions of the two atoms are $ \Gamma_{ii}, ~ i = 1, ~ 2 $ and the excitation is transported from the first atom to the second atom via dipole-dipole interaction. Generic iso-frequency contours of the (b) reciprocal and (c) nonreciprocal photonic media, respectively. (d) The table showing how the upper bound of $ \Gamma_{i \neq j}/\Gamma_{i = j} $ varies from reciprocal to nonreciprocal interaction.}
		\label{Fig1}
	\end{center}
\end{figure*}

Assuming two identical atoms, i.e. $\Gamma_{11} = \Gamma_{22}$, to be initially prepared in the state $ \left | e_1,g_2; 0 \right > $ and interacting via a reciprocal medium $\mathbb{G}(\textbf{r}_1, \textbf{r}_2) = \mathbb{G}(\textbf{r}_2, \textbf{r}_1)$, see Fig. \ref{Fig1}(a), it can be shown that the probabilities of finding them in the excited state are (see Appendix \ref{AppB}):

\begin{equation}\label{Recip}
\begin{split}
	& P_\mathrm{1,e} = \frac{1}{4} \left[  e^{-(\Gamma_{11} + \Gamma_{12})t}  + e^{-(\Gamma_{11} - \Gamma_{12})t} \right]  + \frac{e^{-\Gamma_{11} t}}{2}  \cos   \left(2g_{12}t\right),    \\
    & P_\mathrm{2,e} = \frac{1}{4} \left[  e^{-(\Gamma_{11} + \Gamma_{12})t}  + e^{-(\Gamma_{11} - \Gamma_{12})t} \right]  - \frac{e^{-\Gamma_{11} t}}{2}  \cos   \left(2g_{12}t\right).
\end{split}
\end{equation}

The importance of $ \Gamma_{ij} $ and $ g_{ij} $ can be understood by inspecting above probability equations: (i) $ g_{ij} $ which is related to the real part of the dipole-dipole potential strength is responsible for photon recycling between the atoms. When $ g_{ij} $ is large, the sinusoidal terms cause oscillations in the probabilities related to photons being recycled between the two atoms through the reciprocal medium (Rabi oscillations). However, the strength of photon bouncing back and forth decays with a rate equal to the spontaneous emission. As explained before, this is an evidence that the spontaneous emission competes with the quantum energy transport process. (ii) The larger the ratio $ \Gamma_{12}/\Gamma_{11} $, the higher the probability of photon transport from one emitter to another. However, (iii) there is a fundamental bound such that the physical decay rates given by $ \Gamma_{11} \pm \Gamma_{12} $ must be always non-negative, , otherwise the probability of finding the atoms in the excited state diverges. It implies that the bound $\left | \Gamma_{i \neq j} \right | / \Gamma_{ii} \leq 1 $ must be respected. We call this reciprocity limit (R-limit).

The R-limit means for identical atoms interacting in reciprocal systems $ \mathrm{Im}\mathbb{G}(\textbf{r}_i, \textbf{r}_j) \leq \mathrm{Im}\mathbb{G}(\textbf{r}_i, \textbf{r}_i)  $. This implies that the emitter position is a global maximum of the imaginary part of the Green's function, or, $ \partial_j \mathrm{Im}\mathrm{G}(r=\mathbf{r}^\prime) = 0, ~ j=x, ~y, ~ z $. Breaking this bound leads to higher efficiency for excitation transport, however this is not possible as long as the emitter sits on the Green's function extermum. To break the bound we must provide a situation such that the Green's function spatial derivative becomes nonzero, $ \partial_j \mathrm{G} \neq 0$, at the emitter location. In this situation the Green's function maximum(s) occurs in locations farther from the emitter $ \mathbf{r}_i \neq \mathbf{r}_j $, providing the opportunity of having $ \Gamma_{i \neq j}/\Gamma_{ii} > 1 $. The work \cite{Sorensen_2}, studied a system consisting of TL emitters communicating via  reciprocal surface plasmon modes of a nanowire. The results clearly show that for such a system the R-limit is always respected, but neglected that this is a direct consequence of reciprocity.

In reciprocal environment, this is not possible to have nonzero Green's function spatial derivative at the source position unless there is a large and very sharp physical discontinuity in the system. Even in this situation $ \Gamma_{12} $ barely exceeds the spontaneous emission over a short spatial range, see the discussion in Appendix \ref{AppC}. However as the details are provided in Appendix \ref{AppC}, breaking reciprocity provides a systematic
mechanism to significantly overcome this limit over long-range inter-atom spacing. This can be understood from the following equation which connects the collective decay rate (corresponding to the imaginary part of the potential strength $M_\mathrm{fi}$) to the iso-frequency contours, whose expression over 2D momentum space is written as~\cite{Ying2019,Ying2022}
\begin{equation}\label{Toy_model}
	\Gamma_{12} \propto \int_{\ell_{\omega_0(\mathbf{k})}}     \frac{(\boldsymbol{\mu}_1\cdot \mathbf{e}_\mathbf{k}) (\boldsymbol{\mu}_2\cdot \mathbf{e}_\mathbf{k})^\ast}{v_\mathbf{k}}  e^{i\mathbf{k}\cdot \mathbf{R} } d\ell_\mathbf{k},
\end{equation}
where $\mathbf{e}_\mathbf{k}~(v_\mathbf{k})$ is the electric field (group velocity) of mode $\mathbf{k}$, and  $\mathbf{R}$ is the inter-atomic spacing and $\ell_{\omega_0(\mathbf{k})}$ is the contour with frequency of $\omega_0$ (iso-frequency contour). For the reciprocal case, the weight of each mode $\mathbf{k}$ is uniformly distributed over the iso-frequency contour ($ v_\mathbf{k} $ is a constant and thus it is symmetric in momentum space). This isotropic pattern in momentum space leads to a symmetrical collective decay rate and coherent coupling, i.e. $\Gamma_{12}=\Gamma_{21}$ and $g_{12}=g_{21}$. It also leads to the $R^{-1/2}$ scaling for both $\Gamma_{12}(R)$ and $g_{12}(R)$. However, $ v_\mathbf{k} $ becomes asymmetric under nonreciprocity assumption which leads to an asymmetrical iso-frequency pattern in momentum space, see Fig.~\ref{Fig1} (b) and (c). In nonreciprocal case, the mode weight concentrates around a specific region and results in nonreciprocal couplings $\Gamma_{12}\neq\Gamma_{21}$. Also, the inter-distance scaling reduces to $R^\alpha$, with  $-1/2<\alpha<0$ (longer range interaction). The scaling of the coherent coupling follow the collective decay rate in most scenarios since they are connected by the Kramer-Kronig relation~\cite{Ying2019}. When the iso-frequency contour is symmetric in momentum space, the spatial derivative of the Green's function imaginary part becomes zero (emitter sits on the global maximum of $ \Gamma_{ij} $ and releases momentum symmetrically), i.e, R-limit is respected. Therefore, to dislocate the emitter from the $ \Gamma_{ij} $ global maximum and violate the R-limit, the iso-frequency contour must have an asymmetric shape, see the discussion in Appendix \ref{AppC}. This can be achieved by breaking the physical symmetry of a reciprocal system, or as explained above, by breaking electromagnetic reciprocity. However, as discussed later and in Appendix \ref{AppC}, it is the latter that effectively satisfies the condition to overcome the R-limit.

To demonstrate how nonreciprocity enhances photon transport process, we consider the same configuration shown in Fig. \ref{Fig1}(a), but we assume asymmetric photon-mediated interaction, $ \mathbb{G}(\textbf{r}_i , \textbf{r}_j ) \neq \mathbb{G}(\textbf{r}_j , \textbf{r}_i ) $. In this situation the R-limit bound does not hold any more. In order to find the highest bound we imagine a strong nonreciprocal interaction such that $ \mathbb{G}(\textbf{r}_2, \textbf{r}_1 ) \gg \mathbb{G}( \textbf{r}_1, \textbf{r}_2 ) $ i.e., although the first atom sees the second atom, the second atom does not know about the presence of the first atom. Considering the atomic system initially prepared to be in the state $ \left |e_1,g_2:0 \right > $, then the probability of finding the atoms in the excited state takes the following form (see Appendix \ref{AppB}):
\begin{align}\label{Non-Recip}
	& P_\mathrm{1,e} = e^{-\Gamma_{11} t}, ~ P_\mathrm{2,e} = \frac{ \left |  M_\mathrm{fi}  \right |^2 }{\hbar^2} t^2 e^{-\Gamma_{11}t},
\end{align}
where it can be shown that $  \left |  M_\mathrm{fi}  \right |^2/\hbar^2  = \left |  \Gamma_{21}/2 + ig_{21}  \right |^2$. The first atom which is initially excited decays to its ground state simply via spontaneous emission. But the second atom's state receiving the excitation rises up in a more complicated way. It can be shown that the second atom starts from $ P_\mathrm{2,e} = 0 $ due to its initial preparation, but reaches its maximum probability at $ t = 2/\Gamma_{11} $. At this moment if we assume $ P_\mathrm{2,e} \leq 1 $, a new limit can be found:

\begin{equation}
\left | \Gamma_{21}/2 + ig_{21} \right | \leq e\Gamma_{11}/2
\end{equation}
with $e$ being the neper number. This limit can be also expressed in terms of the reservoir Green's function as $ \left | \mathbb{G}(\textbf{r}_1 \neq \textbf{r}_2) \right | \leq e \left| \mathrm{Im} \mathbb{G}(\textbf{r}_1 = \textbf{r}_2) \right | $ which links the maximum value of the Green's function propagator to the imaginary part of the Green's function at the source position. Therefore asymmetric photon-mediated interaction replaces the R-limit with a new bound as $ \left | \Gamma_{i \neq j}/2 + ig_{i \neq j} \right | \leq e\Gamma_{ii}/2 $, which we call it nonreciprocity limit (NR-limit) in the rest of the paper.

The NR-limit has an important implication. It suggests that by breaking reciprocity, it is possible to go beyond the R-limit upper bound and come up with situations in which $\Gamma_{i \neq j}$ exceeds $\Gamma_{ii}$. This is a key ingredient to achieve large and efficient inter-atomic quantum excitation transport, see the table in Fig. \ref{Fig1}(d) which compares the two limits side-by-side. In what follows we study the atom-atom interaction and inter-atomic quantum excitation transport efficiency via a simple planar reciprocal and nonreciprocal plasmonic waveguide.

\section{Numerical Results}

\begin{figure*}[bth!]
	\begin{center}
		\noindent
		\includegraphics[width=1.8\columnwidth]{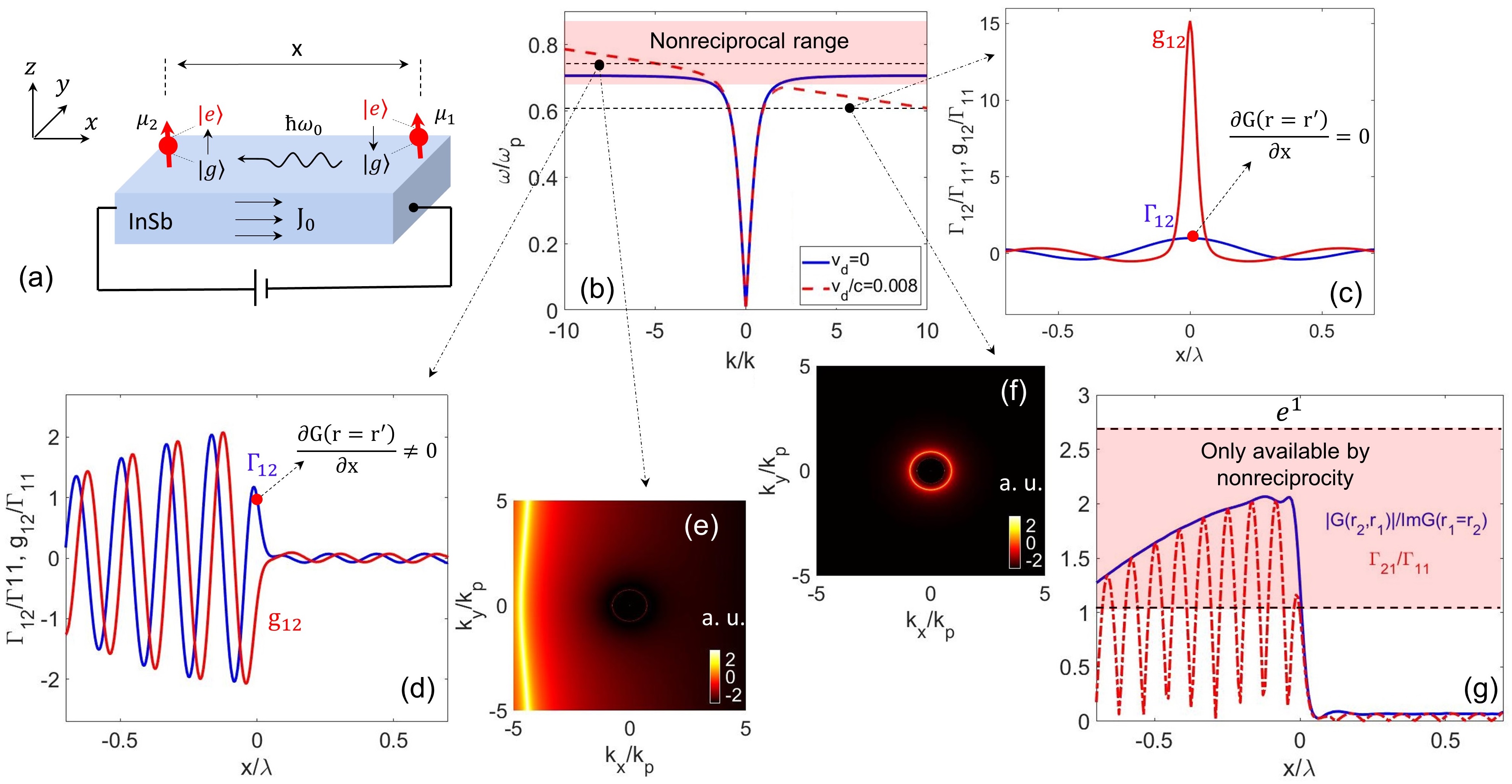}
		\caption{ (a) A plasmonic material (InSb semiconductor) is interfaced with vacuum. Two TL atoms are placed close to the interface in the vacuum region. These atoms dominantly interact via the supported SPPs by the interface. The supported SPPs dispersion line for propagation along $x$-axis is shown in (b) for reciprocal isotropic case, with no current bias, $ v_d = 0 $ and the same configuration but with a DC current with electron drift velocity $ v_d/c = -0.008 $. The red strip demonstrates the frequency range where nonreciprocal SPPs emerge when the InSb is biased. Normalized coupling strength $ g_{12} $ (proportional to the real part of the dipole-dipole potential strength) and decay rate $ \Gamma_{12} $ (proportional to the imaginary part of the dipole-dipole potential strength) as a function of distance between atoms for (c) nonbiased and (d) biased InSb respectively. In reciprocal case $ \omega/ \omega_p = 0.6 $ but for nonreciprocal case $ \omega/ \omega_p = 0.74 $ (a frequency within the nonreciprocal range). The red dot indicates the position of the emitter. As is clear, in nonreciprocal case the Green's function spatial derivative is nonzero. The point source is assumed to be located at $ z_0 = \lambda/40 $ above the interface. $ \lambda $ is the radiation wavelength in free space. The iso-frequency contours for panels (c) and (d) are demonstrated in (e) and (f) respectively. As can be seen the iso-frequency contour is asymmetric/asymmetric in momentum domain for nonbiased/biased plasmnonic region. (d) Normalized total scattered Green's function by its imaginary part and normalized dissipative decay rate by the spontaneous emission for nonreciprocal case, presented in panel (d). The red region where $ \Gamma_{21} > \Gamma_{11} $ becomes available only via nonreciprocity.   }\label{Fig2}
	\end{center}
\end{figure*}

We consider a system as shown in Fig. \ref{Fig2}(a). Two TL atoms are placed on top of an interface between a plasmonic material (InSb semiconductor in our case) and vacuum. The two atoms interact dominantly via the supported surface plasmon-polaritons (SPPs). The supported SPPs propagate at the interface symmetric and omnidirectionally along every angle in $x-y$ plane, establishing a reciprocal interaction between the emitters. For the plasmonic material we employ the Drude model for a free-electron gas with permittivity $\epsilon(\omega) = 1 - \left( \omega / \omega_p \right)^2/( 1 + i\gamma/\omega)$, where $ \omega_p $ is the plasma frequency and $ \gamma $ is the damping rate. The reciprocal dispersion line for SPPs propagating along $x$-axis (the axis connecting the atoms) is shown in Fig. \ref{Fig2}(b), solid blue line. As is known, SPPs exist and propagate reciprocally for frequencies $ \omega \leq \omega_p/\sqrt{2} $ \cite{Novotny}. The calculated dissipative decay rate and coherent coupling terms normalized by spontaneous emission is presented in Fig. \ref{Fig2}(c). The rates are plotted based on the scattered Green's function because it dominate over the vacuum Green's function for our case of interest. In this panel one emitter is located at $x = 0$ (red dot), and the position of the other emitter varies. Note that the slope of Green's function is zero at the emitter position, red dot, i.e., emitter position is the Green's function global maximum, and for any $ |x| > 0 $ the R-limit is respected ($ \Gamma_{12}/ \Gamma_{11} \leq 1$).

The reciprocity of this plasmonic platform can be broken by driving a DC electric current through the substrate with sufficiently high electron drift velocity to tilt the dispersion of surface plasmon-polaritons (SPPs). This method can also be applied to either three-dimensional (3D) conducting materials (metals, degenerately doped semiconductors, and plasmas) \cite{Bliokh,Drifting_electrons} or two-dimensional (2D) media such as graphene \cite{Mario_graphene,Mario_negative_landau}, with a very recent experimental demonstration at optical frequencies \cite{Basov,Efficient_Fizeau}. The effect of drifting electrons on SPPs propagation can be explained in an intuitive way: SPPs are collective charge oscillations coupled to light, hence they are either dragged or opposed by the drifting electrons, which causes surface modes to see different media when propagating along or against the current. The origin of this nonreciprocal behavior is rooted in the frequency Doppler shift due to the electron drift velocity, $ \epsilon (\omega) \rightarrow \epsilon (\omega - \textbf{k} \cdot \textbf{v}_d) $, where $ \textbf{k} $ is the wavevector and $ \textbf{v}_d $ is the electron drifting velocity, see \cite{Drifting_electrons} for further discussion on nonreciprocal SPPs and the Green's function calculation.

The presence of drifting electrons in InSb modifies the SPPs dispersion line as is shown in Fig. \ref{Fig2}(b), dashed red line. The dispersion asymptotic part is tilted as a result of the Doppler shift. Such a tilted dispersion relationship opens up a frequency range supporting nonreciprocal SPPs propagation, see the red region in Fig. \ref{Fig2}(b). Figure \ref{Fig2}(d) shows the normalized dissipative decay rate and coherent coupling terms as a function of emitters spacing at a frequency within the red region (nonreciprocal frequency range) in Fig. \ref{Fig2}(b). The red dot indicates the position of the emitter. It is clear that due to nonreciprocity, the Green's function slope is nonzero at $x = 0$ (emitter location). The iso-frequency contours for these two waveguiding scenarios are demonstrated in Fig. \ref{Fig2} (e) and (f) respectively. As is clear, in reciprocal case, the iso-frequency contour is symmetric in momentum domain which leads to zero spatial slope for Green's function at the emitter location, therefore the R-limit is respected. However, in nonreciprocal case, the iso-frequency contour is asymmetric, therefore the Green's function spatial slope is nonzero at the emitter location. This confirms the discussion in the previous section and the results provided in Appendix \ref{AppC}. Most importantly, we can see that for the nonreciprocal photon-mediated interaction, the R-limit is strongly violated and thus we have $ \Gamma_{12} > \Gamma_{11} $, as predicted by the NR-limit in our analytical discussion. Figure \ref{Fig2}(g) shows the total scattered Green's function normalized by its imaginary part (NR-limit expression) and dissipative decay rate normalized by the spontaneous emission rate for the nonreciprocal case in Fig. \ref{Fig2}(d). It can be seen that: (i) the normalized total Green's function is bounded by $e^1$ from the top and (ii) there are locations far from the emitter where dissipative emission rate exceeds the spontaneous emission rate, all consistent with the NR-limit; see that portion of dashed red and solid blue lines that enter the read area. Having such a large total scattered Green's function and dissipative decay rate is a must for enhancing photon transport efficiency between emitters. Next, we show this enhancement by calculating the probability of finding the atoms in their excited state and also the quantum excitation transport efficiency.

\subsection{Probability of finding the emitters in the excited state}

\begin{figure*}[bth!]
	\begin{center}
		\noindent
		\includegraphics[width=1.5\columnwidth]{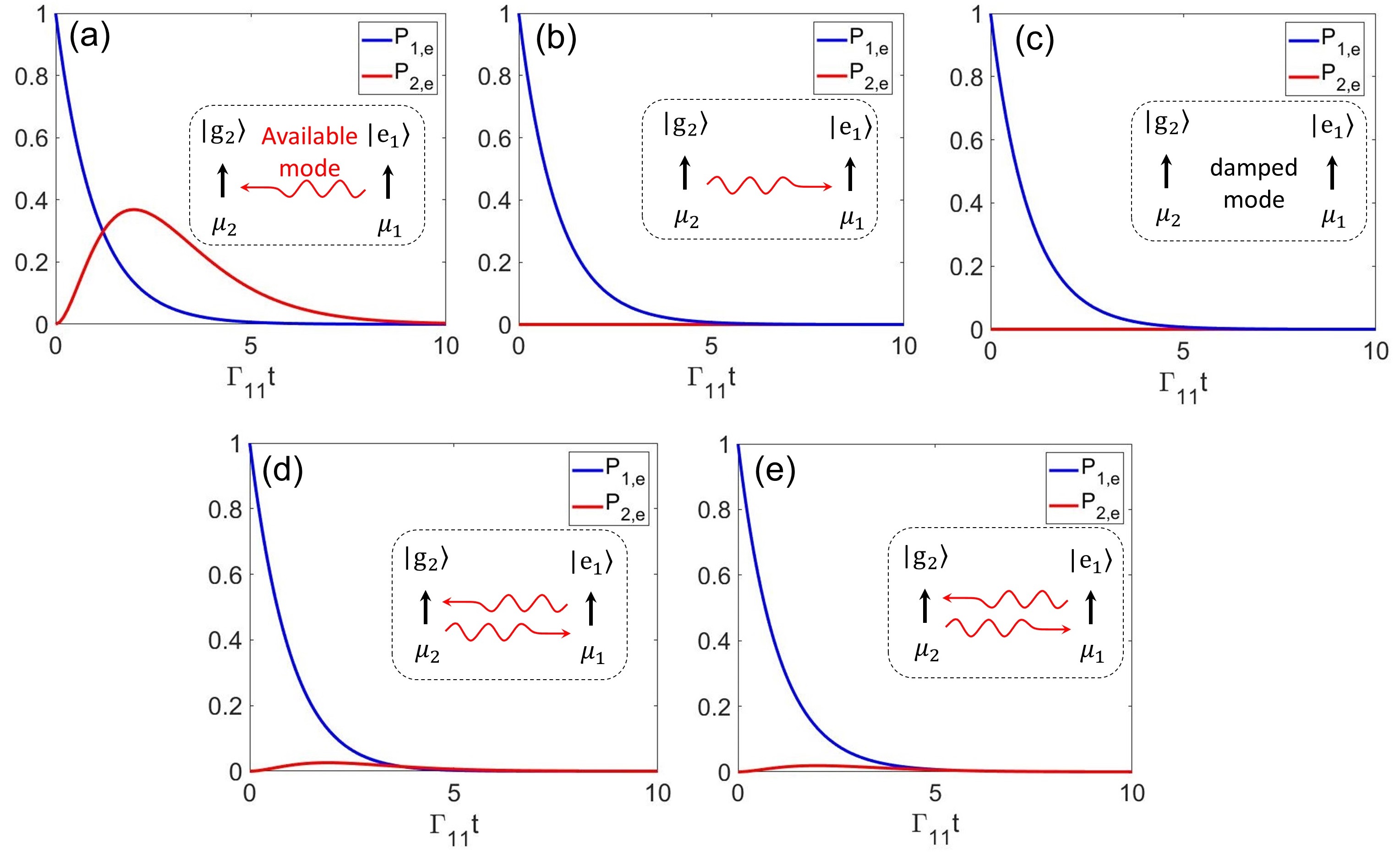}
		\caption{Probability of finding the first and second atom in the excited state as a function time for the atoms to be initially prepared in the state $ \left | e_1,g_2; 0 \right > $ for (a) $ v_d/c = -0.008 $, SPPs propagates from atom 1 to 2, (b) $ v_d/c = 0.008$, SPPs propagate from atom 2 to 1, (c) $v_d = 0$, no propagating mode available for photon exchange, (d) vacuum and (e) $v_d = 0$ at $ \omega/\omega_p = 0.6 $ where the reciprocal modes are available for interaction. The inset in each panel represents the two atoms, their initial state preparation, and available modes to mediate the interaction. }\label{Fig3}.
	\end{center}
\end{figure*}

Let us consider two identical TL emitters placed above the plasmonic waveguide (see Fig. \ref{Fig2}(a)), close to the interface, at $ z_0 = \lambda /40 $ in the vacuum region with spacing $ x = \lambda/2 $ where $\lambda = 2\pi c/\omega_0$ is the radiation wavelength in the free space. The two-emitter system is initially prepared in the state $ | e_1,g_2; 0 \rangle $, such that the right emitter is initially in the excited state while the left one is in the ground state. The evolution of the system can be described by Eq. \ref{ME}.  We first calculate the probability of finding each emitter in the its excited state as the time goes by. Figure \ref{Fig3}(a-c) shows the probabilities when the two emitters interact at a frequency within the nonreciprocal region, red area in Fig. \ref{Fig2}(b) ($ \omega / \omega_p = 0.74 $). Different interaction scenarios are presented in these panels: when the electron stream is toward left, right and when it is turned off. For each interaction scenario the inset represents the two emitters, their initial preparation and the available modes for interaction.

Figure \ref{Fig3}(a) is for the case in which the plasmonic material is driven with electrons with drift velocity $ v_d/c = -0.008 $. In this case, the two emitters interact via a nonreciprocal mode propagating toward left, as shown in Fig. \ref{Fig2}(d). As depicted in the inset of Fig. \ref{Fig3}(a), the supported mode asymmetrically propagates from the initially excited emitter but does not come back. Therefore the excitation from the first emitter can be transported to the second one. It is clear that in this interaction scenario the likelihood of finding the second atom in its excited state is large. In Fig. \ref{Fig3}(b), the electron drifting velocity is reversed, therefore the photonic mode propagates from left to right. As a result, this mode cannot transport the excitation from the initially excited atom (on the right) to the one initially in the ground state (on the left). Consequently, the probability of finding the second emitter in its excited state never rises up and that of the initially excited emitter exponentially decays to zero. Figure \ref{Fig3}(c) shows the case that there is no drifting electron. In this situation there is no propagating mode to establish the interaction between the emitters. Therefore no excitation reaches the second atom while the first atom loses its excitation in the form of heat in the system (releasing the photon into the damped mode). Figure \ref{Fig3}(d) and (e) show the cases that the emitters interact via vacuum and reciprocal plasmonic waveguide (no current in InSb at frequency $ \omega / \omega_p = 0.6 $, see Fig. \ref{Fig2}c) respectively. In these interaction schemes the supported modes are reciprocal. As can be seen the probability of finding the second atom in the excited state is very low. The poor performance of the reciprocal interaction in these panels compared to panel (a) can be understood by the R-limit: $ \Gamma_{12} $ or $ \Gamma_{21} $ cannot exceed $ \Gamma_{11} = \Gamma_{22} $. However, by breaking the reciprocity it becomes possible to have $ \Gamma_{21} $ larger than $ \Gamma_{11} = \Gamma_{22} $ which is essential to have strong emitter-emitter interaction, see Fig. \ref{Fig2}(d) which clearly shows for nonreciprocal interaction the red regions with $ \Gamma_{i \neq j} > \Gamma_{ii} $ is accessible.

\begin{figure}[bth!]
	\begin{center}
		\noindent
		\includegraphics[width=\columnwidth]{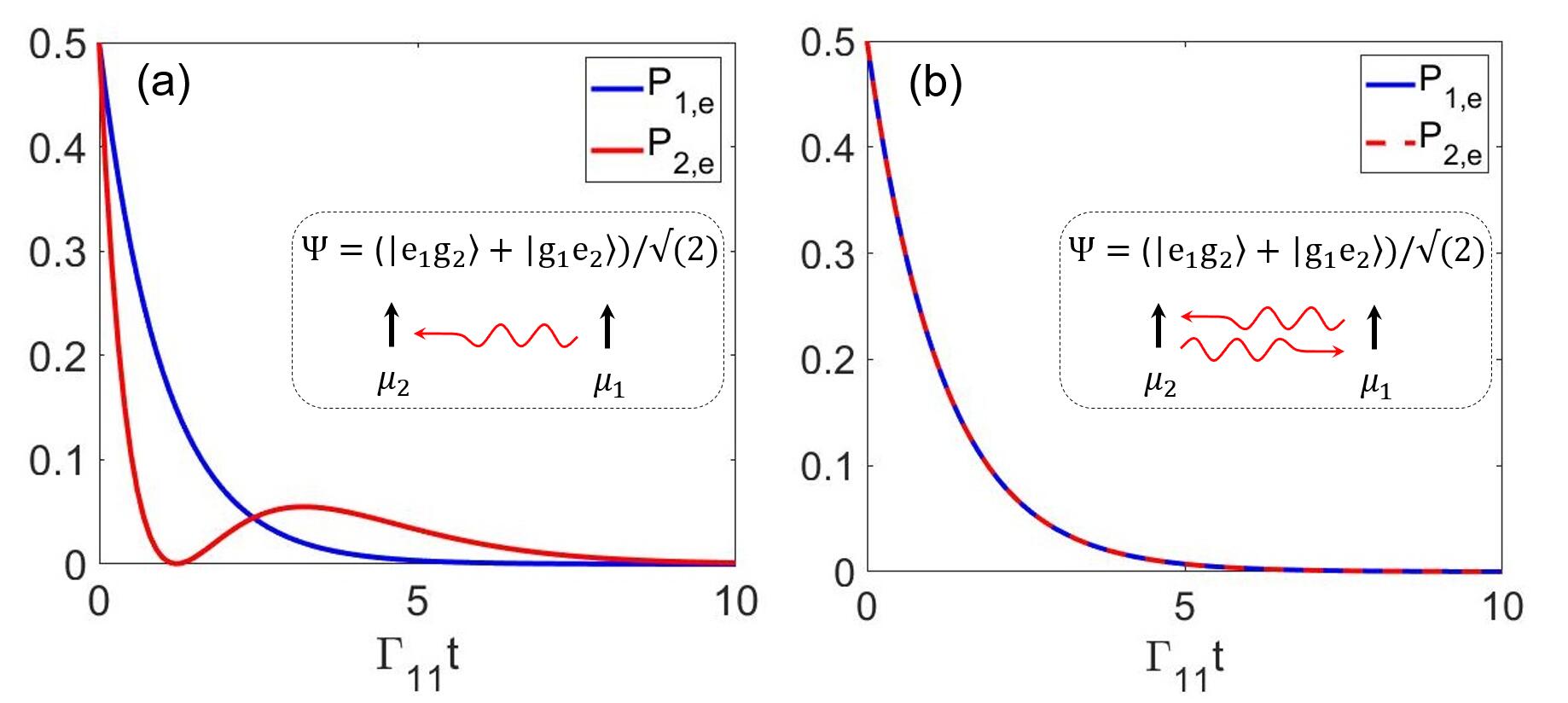}
		\caption{Probability of finding the first and second atom in the excited state as a function time for the atoms to be initially prepared in the Bell entangled state $ \left(\left | e_1,g_2 \right > + \left | g_1,e_2 \right >\right)/\sqrt{2}   $ for (a) nonreciprocal case shown in Fig. \ref{Fig2}(d) and (b) vacuum, for emitter spacing $ x = \lambda /2 $.}\label{Fig4}
	\end{center}
\end{figure}

It is also possible to consider different initial states which can give other possible nonreciprocity assisted dynamical evolution. Figure \ref{Fig4} shows the case of the initial state being the maximally entangled Bell state $ \Psi =  \left(  \left | e_1,g_2 \right> +  \left | g_1,e_1 \right> \right)/ \sqrt{2} $. We consider that the emitters interact through nonreciprocal waveguiding scenario, shown in Fig. \ref{Fig2}(d), and vacuum with the same emitter spacing, $ x = \lambda/2 $. Figure \ref{Fig4} shows the time evolution of the probabilities of finding the atoms in their excited states for both cases. In reciprocal interaction (vacuum), Fig. \ref{Fig4}(b), the probability profiles start from $ P_\mathrm{1,e} = P_\mathrm{2,e} = 0.5$ due to the maximum degree of entanglement of the initial Bell state and the two emitters decay to their ground state with the exact same profile which is an indication of symmetric photon exchange between the emitters. In contrast, when the photon exchange process is asymmetric, see Fig. \ref{Fig4}(a), the decay profiles of the two emitters become asymmetric as well. Similar to previous case the probability profiles start from $ P_\mathrm{1,e} = P_\mathrm{2,e} = 0.5$, then as the time goes by, the probabilities decay in a more complicated manner. The probability of the second emitter being in the excited state (solid red line) experiences a sudden death (the two emitters disentangled for a moment), then the probability experiences a rebirth (the two emitters get entangled again) before decaying exponentially to zero at long times. Next we investigate the efficiency of quantum excitation transport.

\subsection{Quantum excitation transport efficiency}

We consider the same setup as the previous section but here we evaluate the the photon-transport efficiency. To do this we solve the master equation in Eq. \ref{ME}, for two different scenarios: (i) pumped scenario in which energy is injected into the first atom and extracted from the second atom, characterized by $ \Gamma_\mathrm{in}, ~ \Gamma_\mathrm{out} \neq 0 $ with the corresponding solution denoted as $ \rho(t) $ and (ii) non-pumped scenario in which the energy is extracted from the second atom while there is no energy injection in the first atom $ \Gamma_\mathrm{in} = 0, ~ \Gamma_\mathrm{out} \neq 0 $, with the corresponding master equation solution $ \rho_0(t) $. In this situation the two-atom system evolves based on its initial state preparation. For a generic solution the energy fluxes of pumping and extraction are

\begin{align}
	&	P = \frac{\Gamma_\mathrm{in}}{2} \mathrm{Tr} \left[  H_\mathrm{sys} \left(  2\sigma_1^{\dagger} \rho(t) \sigma_1 - \sigma_1 \sigma_1^{\dagger} \rho(t) - \rho(t) \sigma_1 \sigma_1^{\dagger}   \right)  \right] \nonumber \\&
	E = -\frac{\Gamma_\mathrm{out}}{2} \mathrm{Tr} \left[  H_\mathrm{sys} \left(  2\sigma_2 \rho(t) \sigma_2^{\dagger} - \sigma_2^{\dagger} \sigma_2 \rho(t) - \rho(t) \sigma_2^{\dagger} \sigma_2   \right)  \right]
\end{align}
then the transport efficiency is defined as \cite{PRL-Hassani, OQS_7,Mauro_EuroPhysics}
 \begin{equation}
 \chi(t) = \frac{  E(\rho(t)) -  E(\rho_0(t))  }{P(\rho(t))}.
 \end{equation}
Based on this definition if $ \chi(t) = 0 $, then the pumped energy into the first atom does not reach the second one, but if $ \chi(t) = 1 $, all injected energy reaches the second atom without loss. Therefore $ \chi(t) $ varies from $0$ (poor efficiency) to $1$ (efficient photon transport).

\begin{figure}[bth!]
	\begin{center}
		\noindent
		\includegraphics[width=0.8\columnwidth]{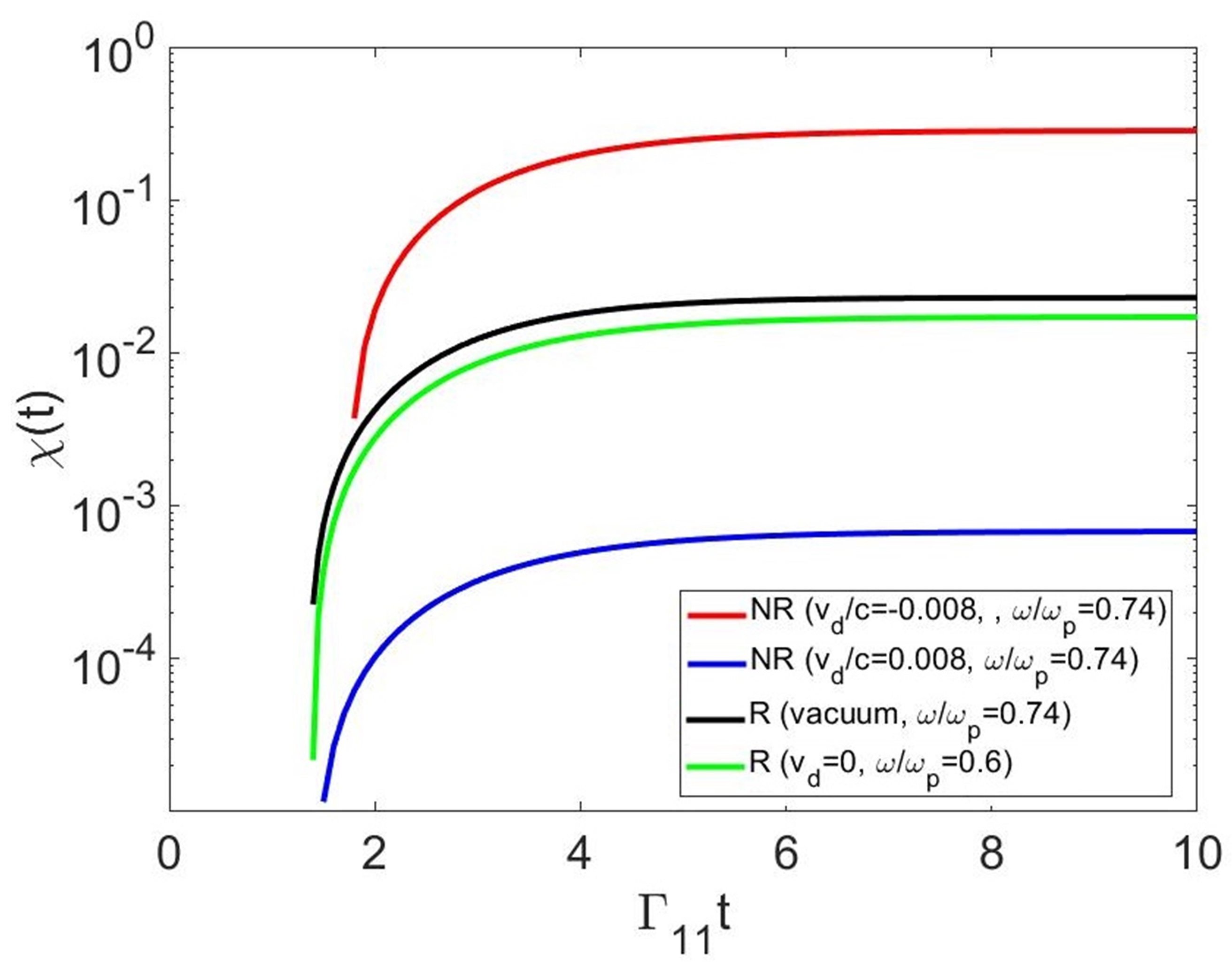}
		\caption{Dynamics of the energy transport efficiency of two interacting atoms for different reciprocal and nonreciprocal environments with spacing $ x = \lambda/2 $ and initial state $ \left | e_1, g_2;0 \right >$ with $ \Gamma_\mathrm{in} = \Gamma_\mathrm{out}  = 0.8 \Gamma_{11}$. The green solid red line corresponds to the biased InSb-vacuum interface with $ v_d/c = -0.008 $, similar to Fig. \ref{Fig2}(d) in which SPPs transport the excitation from the excited atom to the one in its ground state. The solid blue line is for the reversed bias case $ v_d/c = 0.008$ in which the SPP propagation direction is also reversed (SPPs take the excitation away from the atomic system). The solid black and green lines are for two atoms reciprocally interacting in vacuum and nonbiased InSb-vacuum interface respectively.  }\label{Fig5}
	\end{center}
\end{figure}

In the following, we will use the above-mentioned indication to compare transport efficiency between reciprocal and nonreciprocal environments, see Fig. \ref{Fig5}. We evaluate the transport efficiency when the two-atom system is prepared in the initial state $ \left | e_1, g_2; 0 \right> $ and the interaction occurs in different waveguiding scenarios: (i) Nonreciprocal InSb-vacuum interface in which the supported mode delivers the excitation to the second atom initially in the ground state. This situation corresponds to Fig.\ref{Fig2}(d). (ii) The exact same platform but with the reversed biased nonreciprocal InSb. In this case SPPs propagate along the opposite direction (excitation never reaches the second atom initially in the ground state), (ii) Non-biased reciprocal InSb-vacuum interface where the two atoms interact via reciprocal modes, see Fig. \ref{Fig1}(c). For this case the atom transition frequency is lowered down to the region where reciprocal SPPs exist, and finally (iii) vacuum.

For all these photon exchange scenarios, the corresponding probability of finding the two atoms in their excited state are shown in Fig. \ref{Fig3}. Comparing the probability values in this figure suggests that nonreciprocal photon-mediated interaction must lead to a large efficiency in photon transport. The dynamics of $ \chi(t) $ of the two-atom chain in different interaction scenarios are shown in Fig. \ref{Fig5}. It is clear that nonreciprocity leads to orders of magnitude larger steady state transport efficiency $ \chi(t \rightarrow \infty) $. As is discussed in previous sections, such a large efficiency is linked to the large Green's function propagator and dissipative decay rate which becomes available via breaking the symmetry in photon-mediated interaction. Moreover, it can be concluded from Fig. \ref{Fig5} that using the specific plasmonic platforms we employed in this work, varying the electron drift velocity provides the possibility of tuning energy transport, which additionally has the advantage of being an easily accessible parameter by varying an external voltage source.

\section*{Conclusion}

In this work, we have investigated the advantageous properties of nonreciprocity in inter-atomic photon exchange and excitation transport. We have shown that in a system consisting of two TL atoms, breaking electromagnetic reciprocity drastically enhances the probability of photon emission from one atom to another, which in turn enhances the excitation transport process. Our approach is based on the system 3D Green's function and we have found that breaking reciprocity changes the spatial slope of the Green's function at the location of the atom itself from zero to a nonzero value. As a direct consequence, it becomes possible for the cooperative decay rate to exceed the spontaneous emission rate; a condition that cannot be reached for two identical atoms interacting in a reciprocal and translation-invariant platform. Thus, stronger dipole-dipole interaction can be achieved. We have also provided an example where two identical atoms interact via a reciprocal plasmonic 3D platform, i.e., InSb-vacuum interface. The reciprocity of such a system can be violated in a magnet-free manner by driving an electron stream within the InSb material. In the presence of drifting electrons, atoms interact via nonreciprocal SPPs. We have numerically demonstrated a significant enhancement in photon exchange process by calculating the efficiency of quantum excitation transport. In addition, we have discussed the possibility of tuning energy transport in such a platform by varying the electrons drift velocity, which can be an easily accessible parameter by varying an external voltage. Our findings presented in this paper suggest that electromagnetic nonreciprocity can play a key role in the development of quantum technologies requiring efficient and tunable energy transport and energy management at the microscopic scale and single photon level.


\section*{ACKNOWLEDGMENTS}
The authors acknowledge the financial support by NSF through QLCI-CI: Hybrid Quantum Architectures and Networks.

\begin{widetext}
\appendix

\section{Resonant Dipole-Dipole Interaction and Resonant Energy Transfer Rate} \label{AppA}

Considering the electric field operator as
\begin{equation}
	\hat{\textbf{E}}(r, \omega_k) = i \sqrt{ \frac{\hbar}{ \pi \epsilon_0 } } \frac{\omega_k^2}{c^2} \int_{\mathbf{r}^\prime} d\mathbf{r}^\prime \mathbb{G}(r,\mathbf{r}^\prime, \omega_k)  \sqrt{  \mathrm{Im}\epsilon(\mathbf{r}^\prime) } \hat{a}(\mathbf{r}^\prime, \omega_k),
\end{equation}
the interaction Hamiltonian under rotating wave approximation can written as
\begin{equation}
	H_\mathrm{int}  = - \left( \sigma^{\dagger}_1  \boldsymbol{\mu} \cdot \hat{\textbf{E}}(\mathbf{r}_1)   + \sigma_1 \hat{\textbf{E}}^{\dagger}(\mathbf{r}_1) \cdot  \boldsymbol{\mu}  \right) - \left( \sigma^{\dagger}_2  \boldsymbol{\mu} \cdot \hat{\textbf{E}}(\mathbf{r}_2)   + \sigma_2 \hat{\textbf{E}}^{\dagger}(\mathbf{r}_2) \cdot  \boldsymbol{\mu}  \right)
\end{equation}
where the atoms are supposed to have same resonance frequency $ \omega_{0} $, same dipole transition moment $ \boldsymbol{\mu} $ and placed at positions $ \mathbf{r}_1 $ and $ \mathbf{r}_2 $. All dipole-dipole transitions can be calculated from the transition matrix element,

\begin{equation}
	M_\mathrm{fi} = \left <  \mathrm{f} \right | H_\mathrm{int}\left | \mathrm{i} \right > + \sum_m \frac{  \left < \mathrm{f} \right | H_\mathrm{int}\left | \mathrm{m} \right >\left < \mathrm{m} \right | H_\mathrm{int} \left | \mathrm{i} \right >    }{ E_\mathrm{m} - E_\mathrm{i} } + \cdots
\end{equation}

Considering the initial and final states as $ \left | \mathrm{i} \right > = \left | e_1g_2, 0 \right > $ and $ \left | \mathrm{f} \right > = \left | g_1e_2, 0 \right > $ i.e., transferring one photon from the first atom to the second atom, the corresponding transition matrix element can be found as follow,

\begin{equation}
	H_\mathrm{int} \left | \mathrm{i} \right > = i \sqrt{ \frac{ \hbar }{ \pi \epsilon_0 } } \frac{ \omega_k^2 }{c^2} \int_{\mathbf{r}^\prime} d\mathbf{r}^\prime \boldsymbol{\mu} \cdot \mathbb{G}(\mathbf{r}_1, \mathbf{r}^\prime, \omega_k) \sqrt{\mathrm{Im}\epsilon(\mathbf{r}^\prime, \omega_k)}    \left | g_1g_2, 1_{\omega_k, \mathbf{r}^\prime} \right >
\end{equation}
the final state is achieved via a generic intermediate state $   \left |\mathrm{m} \right > =  \left | g_1g_2, 1_{\omega_{k^\prime}, R}  \right > $ with energy $ E_\mathrm{m} = \hbar \omega_{k^\prime} $, therefore,

\begin{equation}
	\left < \mathrm{m} \right | H_\mathrm{int} \left | \mathrm{i} \right > =  i \sqrt{ \frac{ \hbar }{ \pi \epsilon_0 } } \frac{ \omega_k^2 }{c^2} \int_{\mathbf{r}^\prime} d\mathbf{r}^\prime \boldsymbol{\mu} \cdot \mathbb{G}(\mathbf{r}_1, \mathbf{r}^\prime, \omega_k) \sqrt{\mathrm{Im}\epsilon(\mathbf{r}^\prime, \omega_k)}    \left < g_1g_2, 1_{\omega_{k^\prime}, R}   | g_1g_2, 1_{\omega_k, \mathbf{r}^\prime} \right >
\end{equation}

Considering the final and intermediate states as $ \left | \mathrm{f} \right > = \left | g_1e_2, 0 \right > $ and $   \left |\mathrm{m} \right > =  \left | g_1g_2, 1_{\omega_{k^\prime}, R}  \right > $, the term $ 	\left < \mathrm{f} \right | H_\mathrm{int} \left |\mathrm{m} \right >  $ can be found as follow,

\begin{align}
	& H_\mathrm{int} \left | \mathrm{m} \right > = - i \sqrt{ \frac{ \hbar }{ \pi \epsilon_0 } } \frac{ \omega_{k^\prime}^2 }{c^2} \int_{R} dR  \mathbb{G}^{\dagger}(\mathbf{r}_2, R, \omega_{k^\prime}) \cdot \boldsymbol{\mu} \sqrt{\mathrm{Im}\epsilon(R, \omega_{k^\prime})}    \left | g_1e_2, 0 \right >, \nonumber \\ &
	\left < \mathrm{f} \right |H_\mathrm{int} \left | \mathrm{m} \right > =  - i \sqrt{ \frac{ \hbar }{ \pi \epsilon_0 } } \frac{ \omega_{k^\prime}^2 }{c^2} \int_{R} dR  \mathbb{G}^{\dagger}(\mathbf{r}_2, R, \omega_{k^\prime}) \cdot \boldsymbol{\mu} \sqrt{\mathrm{Im}\epsilon(R, \omega_{k^\prime})}
\end{align}
consequently,

\begin{align}
	\frac{ \left < \mathrm{f} \right |H_\mathrm{int} \left | \mathrm{m} \right >  \left < \mathrm{m} \right |H_\mathrm{int} \left | \mathrm{i} \right >    }{ E_\mathrm{i} - E_\mathrm{f} } & = \frac{\hbar}{ \pi \epsilon_0  }  \int_{\omega_k} \int_{\omega_{k^\prime}}  \frac{  d \omega_{k} d\omega_{k^\prime}  }{\hbar \left( \omega_{0} - \omega_{k^\prime} \right)} \frac{\omega_{k}^2 \omega^2_{k^\prime}}{c^4} \int_{\mathbf{r}^\prime} \int_R \boldsymbol{\mu} \cdot \mathbb{G} (\mathbf{r}_1, \mathbf{r}^\prime, \omega_k) \mathbb{G}^{\dagger} (\mathbf{r}_2, R, \omega_{k^\prime}) \cdot \boldsymbol{\mu} \nonumber \\ &
	\times \sqrt{ \mathrm{Im}\epsilon(\mathbf{r}^\prime, \omega_{k}) } \sqrt{ \mathrm{Im}\epsilon(R, \omega_{k}') }  \left < g_1g_2, 1_{\omega_{k^\prime}, R}   | g_1g_2, 1_{\omega_k, \mathbf{r}^\prime} \right >
\end{align}
where $ \left < g_1g_2, 1_{\omega_{k^\prime}, R}   | g_1g_2, 1_{\omega_k, \mathbf{r}^\prime} \right > = \delta(R - \mathbf{r}^\prime) \delta ( \omega_{k} - \omega_{k^\prime} ) $. Because of this orthogonality, above equation reduces to

\begin{equation}\label{M_fi}
	\frac{ \left < \mathrm{f} \right |H_\mathrm{i} \left | \mathrm{m} \right >  \left < \mathrm{m} \right |H_\mathrm{i} \left |\mathrm{i} \right >    }{ E_\mathrm{i} - E_\mathrm{f}  }  = \frac{-1}{\pi \epsilon_0} \int_{\omega_k} \frac{d \omega_{k}}{  \omega_{k} - \omega_{0} } \frac{ \omega_{k}^2}{c^2} \boldsymbol{\mu} \cdot \left[ \frac{ \omega_{k}^2}{c^2} \int_{\mathbf{r}^\prime} d\mathbf{r}^\prime  \mathrm{Im} \epsilon(\mathbf{r}^\prime, \omega_{k})  \mathbb{G} (\mathbf{r}_1, \mathbf{r}^\prime, \omega_k) \mathbb{G}^{\dagger} (\mathbf{r}_2, \mathbf{r}^\prime, \omega_{k})    \right] \cdot \boldsymbol{\mu}
\end{equation}

For a the most general case,

\begin{equation}
	\frac{ \omega_{k}^2}{c^2} \int_{\mathbf{r}^\prime} d\mathbf{r}^\prime  \mathrm{Im} \epsilon(\mathbf{r}^\prime, \omega_{k})  \mathbb{G} (\mathbf{r}_1, \mathbf{r}^\prime, \omega_k) \mathbb{G}^{\dagger} (\mathbf{r}_2, \mathbf{r}^\prime, \omega_{k}) = \left( \mathbb{G}(\mathbf{r}_1, \mathbf{r}_2) - \mathbb{G}^{\dagger}(\mathbf{r}_2, \mathbf{r}_1)  \right)/2i
\end{equation}
for a system with strong nonreciprocity, $ \mathbb{G} (\mathbf{r}_1, \mathbf{r}_2) \neq 0 $, $ \mathbb{G} (\mathbf{r}_2, \mathbf{r}_1) = 0 $,

\begin{align}
	M_\mathrm{fi} = \frac{-1}{\pi \epsilon_0} \int_{\omega_k} \frac{d \omega_{k}}{  \omega_{k} - \omega_{0} } \frac{ \omega_{k}^2}{c^2}   \boldsymbol{\mu} \cdot \left[\frac{ \mathbb{G} (\mathbf{r}_1, \mathbf{r}_2, \omega_k) }{2i}\right] \cdot \boldsymbol{\mu}
\end{align}
where utilizing the following relationship,

\begin{align}\label{identity}
	& \int \frac{f(z)}{z -z_0} dz = i\pi f(z_0) + \mathcal{PV} \int \frac{f(z)}{z -z_0} dz \nonumber \\&
	\mathcal{PV} \int \frac{\mathrm{Re}f(z)}{z -z_0} dz = -\pi \mathrm{Im}f(z_0) \nonumber \\ &
	\mathcal{PV} \int \frac{\mathrm{Im}f(z)}{z -z_0} dz = \pi \mathrm{Re}f(z_0)
\end{align}
we get

\begin{equation}
	M_\mathrm{fi} = \frac{-\omega_{0}^2}{\epsilon_0 c^2} \boldsymbol{\mu} \cdot \mathbb{G} (\mathbf{r}_1, \mathbf{r}_2, \omega_0)  \cdot \boldsymbol{\mu}
\end{equation}
decomposing the Green's function into real and imaginary parts, gives $ 	M_\mathrm{fi} = -\hbar \left(g_{12} + i\Gamma_{12}/2\right) $ where

\begin{align}
	& \Gamma_{12} = \frac{2 \omega_0^2}{\hbar \epsilon_0 c^2} \boldsymbol{\mu} \cdot \mathrm{Im}\mathbb{G} (\mathbf{r}_1, \mathbf{r}_2, \omega_0)  \cdot \boldsymbol{\mu} \nonumber \\ &
	g_{12} = \frac{ \omega_0^2}{\hbar \epsilon_0 c^2} \boldsymbol{\mu} \cdot \mathrm{Re}\mathbb{G} (\mathbf{r}_1, \mathbf{r}_2, \omega_0)  \cdot \boldsymbol{\mu}
\end{align}

The resonance energy transfer rate between state $ \left | e_1, g_2, 0 \right > $ and $ \left | g_1, e_2, 0 \right > $ can be calculated using the Fermi’s Golden rule,

\begin{equation}
	\Gamma_{\mathrm{i} \rightarrow \mathrm{f} } = \Gamma_{12} =  \frac{2\pi}{\hbar^2 } \left | M_\mathrm{fi} \right |^2 = \frac{2\pi}{\hbar^2} \frac{\omega^4}{ \epsilon_0^2 c^4 } \left |   \boldsymbol{\mu} \cdot \mathbb{G} (\mathbf{r}_1, \mathbf{r}_2, \omega_0)  \cdot \boldsymbol{\mu}  \right |^2
\end{equation}

Above equation means if the system is prepared in the state  $ \left | e_1, g_2, 0 \right > $, then the energy is transferred to the second atom with the above rate, but if the initial preparation is  $ \left | g_1, e_2, 0 \right > $, then there is no energy transfer from the second emitter toward the first one. For system respecting reciprocity the energy transfer rate has the above form but energy transfer process is bilateral.

\section{Symmetric and Asymmetric Photon Exchange Between TL Systems}\label{AppB}

The master equation valid for reciprocal and nonreciprocal environments is \cite{PRA-Hassani,PRL-Hassani}:

\begin{align}
	\frac{\partial \rho }{\partial t} & = \frac{-i}{\hbar} \left[ H_\mathrm{sys}, \rho(t)  \right] + \sum_{i =1}^{N} \frac{ \Gamma_{ii} }{ 2} \left(  2\sigma_i \rho(t) \sigma_i^{\dagger} -  \sigma_i^{\dagger} \sigma_i \rho(t) -  \rho(t) \sigma_i^{\dagger} \sigma_i   \right) \nonumber \\&
	+ \sum_{i,j}^{i \neq j} \frac{ \Gamma_{ij} }{2 } \left( \left[  \sigma_j \rho(t), \sigma_i^{\dagger} \right] + \left[  \sigma_i, \rho(t)\sigma_j^{\dagger} \right]   \right) + \sum_{i,j}^{i \neq j} g_{ij} \left( \left[  \sigma_j \rho(t), -i \sigma_i^{\dagger} \right] + \left[  i \sigma_i, \rho(t)\sigma_j^{\dagger} \right]   \right) \nonumber \\&
	+ \frac{\Gamma_\mathrm{in}}{2} \left(  2\sigma_1^{\dagger} \rho(t) \sigma_1 -  \sigma_1 \sigma_1^{\dagger} \rho - \rho \sigma_1 \sigma_1^{\dagger} \right) + \frac{\Gamma_\mathrm{out}}{2} \left(  2\sigma_N \rho(t) \sigma_N^{\dagger} -  \sigma_N^{\dagger} \sigma_N \rho - \rho \sigma_N^{\dagger} \sigma_N \right)
\end{align}
where
\begin{align}
	& \Gamma_{ij} = \frac{2\omega_0^2}{\epsilon_0 \hbar c^2}  \boldsymbol{\mu}_{ i} \cdot \mathrm{Im} \left[ \mathbb{G} (\textbf{r}_i, \textbf{r}_j, \omega_{0}) \right] \cdot \boldsymbol{\mu}_{ j} \nonumber \\&
	g_{ij} = \frac{\omega_0^2}{\epsilon_0 \hbar c^2}  \boldsymbol{\mu}_{ i} \cdot \mathrm{Re} \left[ \mathbb{G} (\textbf{r}_i, \textbf{r}_j, \omega_{0}) \right] \cdot \boldsymbol{\mu}_{ j}
\end{align}
are the dissipative decay rate and the coherent coupling terms, $ \omega_{0} $ is the TL system transition frequency, $ \boldsymbol{\mu} $ is the dipole moment operator, $ \sigma_i/ \sigma_i^{\dagger} $ are the atomic energy lowering/raising operators and $ \Gamma_\mathrm{in} / \Gamma_\mathrm{out}$ are the rates of energy pumping and extraction from the first and N$th$ TL system. Assuming only two emitters in the environment, the basis can be defined as follow
\begin{align}\label{Eq:basis}
	&\left|1\right> = \left|g_1\right> \otimes \left|g_2\right> = \left|g_1,g_2\right>,~ \left|2\right> = \left|e_1\right> \otimes \left|e_2\right> = \left|e_1,e_2\right> \notag \\ & \left|3\right> = \left|g_1\right> \otimes \left|e_2\right> = \left|g_1,e_2\right>,~ \left|4\right> = \left|e_1\right> \otimes \left|g_2\right> = \left|e_1,g_2\right>.
\end{align}
it can be shown that operators act on these basis vectors as\\
\newline
\begin{tabular}{|c|c|c|c|c|}
	\hline                              & $\left|1\right>$        & $\left|2\right>$         & $\left|3\right>$           &  $\left|4\right>$ \\
	\hline $\hat{\sigma}_1$             & 0  & $\left|3\right>$ & 0  &  $\left|1\right>$ \\
	\hline $\hat{\sigma}_1^{\dagger}$   &  $\left|4\right>$  & 0  & $\left|2\right>$  &  0  \\
	\hline $\hat{\sigma}_2$    & 0  & $\left|4\right>$   & $\left|1\right>$  &  0\\
	\hline $\hat{\sigma}_2^{\dagger}$   & $\left|3\right>$  &  0  & 0  & $\left|2\right>$ \\
	\hline
\end{tabular}
~~
\begin{tabular}{|c|c|c|c|c|}
	\hline                              & $\left<1\right|$        & $\left<2\right|$         & $\left<3\right|$           &  $\left<4\right|$ \\
	\hline $\hat{\sigma}_1$             & $\left<4\right|$   &  0  & $\left<2\right|$  &  0 \\
	\hline $\hat{\sigma}_1^{\dagger}$   & 0 & $\left<3\right|$   & 0  &  $\left<1\right|$ \\
	\hline $\hat{\sigma}_2$             &  $\left<3\right|$ & 0 & 0 & $\left<2\right|$ \\
	\hline $\hat{\sigma}_2^{\dagger}$   & 0  & $\left<4\right|$   & $\left<1\right|$  &  0 \\
	\hline
\end{tabular}
~~
\begin{tabular}{|c|c|c|c|c|}
	\hline                                                 & $\left|1\right>$        & $\left|2\right>$        & $\left|3\right>$                 &  $\left|4\right>$ \\
	\hline $\hat{\sigma}_1^{\dagger}\hat{\sigma}_1$        & 0  & $\left|2\right>$   & 0  &  $\left|4\right>$ \\
	\hline $\hat{\sigma}_1^{\dagger}\hat{\sigma}_2$        &  0   & 0 & $\left|4\right>$ &  0  \\
	\hline $\hat{\sigma}_2^{\dagger}\hat{\sigma}_1$   &  0  & 0  & 0  &  $\left|3\right>$\\
	\hline $\hat{\sigma}_2^{\dagger}\hat{\sigma}_2$        & 0   & $\left|2\right>$  & $\left|3\right>$ &  0  \\
	\hline
\end{tabular}
~~
\begin{tabular}{|c|c|c|c|c|}
	\hline                                                 & $\left<1\right|$        & $\left<2\right|$         & $\left<3\right|$           &  $\left<4\right|$ \\
	\hline $\hat{\sigma}_1^{\dagger}\hat{\sigma}_1$        & 0  & $\left<2\right|$   & 0  &  $\left<4\right|$ \\
	\hline $\hat{\sigma}_1^{\dagger}\hat{\sigma}_2$        & 0                       & 0                        & 0  &  $\left<3\right|$  \\
	\hline $\hat{\sigma}_2^{\dagger}\hat{\sigma}_1$        & 0                       & 0                        & $\left<4\right|$  & 0\\
	\hline $\hat{\sigma}_2^{\dagger}\hat{\sigma}_2$        & 0 & $\left<2\right|$  & $\left<3\right|$  &  0 \\
	\hline
\end{tabular} \\\\
and assuming a non-symmetric dyadic Green's function, $ \mathbb{G}(\textbf{r}_i, \textbf{r}_j ) \neq \mathbb{G}( \textbf{r}_j, \textbf{r}_i ) $, we get the following set of coupled differential equation describing the dynamics of the density matrix elements,

\begin{align}
	& \frac{\partial \rho_{11}}{\partial t} = \Gamma_{11} \rho_{44} + \left( \Gamma_{11} + \Gamma_\mathrm{out} \right) \rho_{33} + \gamma\rho_{34} + \gamma^*\rho_{43} - \Gamma_\mathrm{in} \rho_{11} +\nu \rho_{43} + \nu^* \rho_{34} \nonumber \\&
	\frac{\partial \rho_{12}}{ \partial t } = -\frac{\rho_{12}}{2} \left( 2\Gamma_{11} + \Gamma_\mathrm{out} + \Gamma_\mathrm{in}  \right) \nonumber \\&
	\frac{\rho_{13}}{ \partial t } = \Gamma_{11} \rho_{42} - \frac{\rho_{13}}{2} \left(  \Gamma_{11} + \Gamma_\mathrm{out} \right) + \gamma \left( \rho_{32} - \rho_{14} \right) - \Gamma_\mathrm{in} \rho_{13} + \nu^* \rho_{32} \nonumber \\&
	\frac{\rho_{14}}{ \partial t } = -\frac{\Gamma_{11}}{2} \rho_{14} + \rho_{32} \left(  \Gamma_{11} + \Gamma_\mathrm{out} \right) + \gamma^* \rho_{42} - \frac{\Gamma_\mathrm{in}}{2} \rho_{14} + \nu \left(  \rho_{42} - \rho_{13} \right)
\end{align}

\begin{align}
	& \frac{\partial \rho_{21}}{\partial t} = - \frac{\rho_{21}}{2} \left(  2 \Gamma_{11} + \Gamma_\mathrm{out} + \Gamma_\mathrm{in} \right) \nonumber \\ &
	\frac{\partial \rho_{22}}{\partial t} = -\rho_{22} \left( 2\Gamma_{11} + \Gamma_\mathrm{out} \right) + \Gamma_\mathrm{in} \rho_{33} \nonumber \\&
	\frac{\partial \rho_{23}}{\partial t} = -\rho_{23} \left(  \Gamma_{11} + \Gamma_\mathrm{in} + \Gamma_\mathrm{out} \right) - \gamma \rho_{24} \nonumber \\&
	\frac{\partial \rho_{24}}{\partial t} = -\rho_{24} \left( 1.5\Gamma_{11} + \frac{\Gamma_\mathrm{out}}{2} \right) + \Gamma_\mathrm{in} \rho_{31} - \nu \rho_{23}
\end{align}

\begin{align}
	& \frac{\partial \rho_{31}}{\partial t} =  \Gamma_{11} \rho_{24} - \frac{\rho_{31}}{2} \left( \Gamma_{11} + \Gamma_\mathrm{out} \right) + \gamma^* \left( \rho_{23} - \rho_{41} \right) - \Gamma_\mathrm{in} \rho_{31} + \nu \rho_{23} \nonumber \\&
	\frac{\partial \rho_{32}}{\partial t} = -\rho_{32} \left( 1.5\Gamma_{11} + \Gamma_\mathrm{out}  + \frac{\Gamma_\mathrm{in}}{2}    \right) - \gamma^* \rho_{42} \nonumber \\&
	\frac{\partial \rho_{33}}{\partial t} = \Gamma_{11} \rho_{22} - \rho_{33} \left( \Gamma_{11} + \Gamma_\mathrm{out} + \Gamma_\mathrm{in} \right) - \gamma \rho_{34} - \gamma^* \rho_{43} \nonumber \\&
	\frac{\partial \rho_{34}}{\partial t} = - \frac{\rho_{34}}{2} \left(  2\Gamma_{11} + \Gamma_\mathrm{out} + \Gamma_\mathrm{in} \right)  + \gamma^* \left( \rho_{22} - \rho_{44} \right) + \nu \left( \rho_{22} - \rho_{33} \right)
\end{align}

\begin{align}
	& \frac{\partial \rho_{41}}{\partial t} =  - \frac{\Gamma_{11} }{2} \rho_{41} + \rho_{23} \left( \Gamma_{11} + \Gamma_\mathrm{out} \right) + \gamma \rho_{24} - \frac{\Gamma_\mathrm{in}}{2}\rho_{41} + \nu^* \left( \rho_{24} - \rho_{31} \right) \nonumber \\&
	\frac{\partial \rho_{42}}{\partial t} = -\frac{\rho_{42}}{2} \left( 3\Gamma_{11} + \Gamma_\mathrm{out} \right) + \Gamma_\mathrm{in} \rho_{13} -\nu^* \rho_{32} \nonumber \\&
	\frac{\partial \rho_{43}}{\partial t} = -\frac{\rho_{43}}{2} \left(  2\Gamma_{11} + \Gamma_\mathrm{out} + \Gamma_\mathrm{in} \right) + \gamma\left( \rho_{22} - \rho_{44} \right) + \nu^* \left(  \rho_{22} - \rho_{33} \right) \nonumber \\&
	\frac{\partial \rho_{44}}{\partial t} = -\Gamma_{11} \rho_{44} + \rho_{22}\left(  \Gamma_{11} + \Gamma_\mathrm{out} \right) + \Gamma_\mathrm{in} \rho_{11} -\nu \rho_{43} - \nu^* \rho_{34}
\end{align}
where $ \gamma = \Gamma_{21}/2 + ig_{21} $ and $ \nu = \Gamma_{12}/2 + ig_{12} $. Let us assume that the system is initially prepared in the state $ \left | 4 \right > = \left | e_1, g_2 \right > $. In no pumping/extraction scenario $ \Gamma_\mathrm{in} = \Gamma_\mathrm{out} = 0 $, by solving above system of coupled differential equation, it can be shown that the probabilities of finding the emitters in their excited state are
\begin{align}
	& P_\mathrm{1,e} = \frac{1}{4} \left[  e^{-(\Gamma_{11} + \Gamma_{12})t}  + e^{-(\Gamma_{11} - \Gamma_{12})t} \right]  + \frac{e^{-\Gamma_{11} t}}{2}  \cos   \left(2g_{12}t\right) \nonumber \\&
	P_\mathrm{2,e} = \frac{1}{4} \left[  e^{-(\Gamma_{11} + \Gamma_{12})t}  + e^{-(\Gamma_{11} - \Gamma_{12})t} \right]  - \frac{e^{-\Gamma_{11} t}}{2}  \cos   \left(2g_{12}t\right)
\end{align}
for the reciprocal environment and
\begin{align}\label{Non-Recip}
	& P_\mathrm{1,e} = e^{-\Gamma_{11} t}, ~ P_\mathrm{2,e} = \left |  \frac{\Gamma_{21}}{2} + ig_{21}  \right |^2 t^2 e^{-\Gamma_{11}t}
\end{align}
for the extreme nonreciprocal case $ \mathbb{G}(\textbf{r}_2, \textbf{r}_1 ) \gg \mathbb{G}( \textbf{r}_1, \textbf{r}_2 ) $.

\section{Nonreciprocity and Nonzero Green's Function Slope at the Emitter Location } \label{AppC}

Here we discuss how electromagnetic nonreciprcoty leads to nonzero Green's function slope at the emitter location. To understand this, we examine the Green's function of an emitter radiating near a planar and transitionally invariant system, as shown in Fig. \ref{Fig1}(a) of the main text. The dyadic Green’s function angular spectrum representation takes the following form \cite{Novotny},

\begin{equation}
	\mathbb{G}(\textbf{r}_1, \textbf{r}_2) = \frac{i}{8 \pi^2} \int_{k_x}\int_{k_y} \mathbb{M}(k_x,k_y) e^{i(k_{x}(x_1 - x_2) + k_y(y_1 - y_2) + k_z(z_1 - z_2)) } dk_x dk_y
\end{equation}
where $ k_z = \sqrt{ \left(\omega/c\right)^2 - k_x^2 - k_y^2 } $ with $ k_x, ~ k_y $ being the in-plane wavenumbers and $ \mathbb{M} $ is a $ 3\times 3 $ matrix that connects the $ mth $ component of the generated electric field to a $n$-polarized emitter, with $m,~n$ being the coordinate directions. Assuming two identical and vertically-polarized emitters located at a distance $ \textbf{R } = \textbf{r}_2 - \textbf{r}_1$ from each other, then the dissipative decay rate as function of their spacing is found to be:

\begin{align}
	 \Gamma_{12} (R) = \frac{ 2 \mu_{1,z}^2 \omega_0^2 }{\epsilon_0 \hbar c^2} \mathrm{Im}\left[ \frac{i}{8 \pi^2} \int_{k_x} \int_{k_y} \mathrm{M}_{zz}(k_x,k_y) e^{i(k_{x}(x_1 - x_2) + k_y(y_1 - y_2) + k_z(z_1 - z_2)) } dk_x dk_y  \right]
\end{align}
where $ \mathrm{M}_{zz} $ is replaced instead of $ \mathbb{M} $. The interaction reciprocity or nonreciprocity is encoded in $\mathrm{M}_{zz}$. At a fixed frequency $ \omega_{0} $, the poles of $ \mathrm{M}_{zz} $ form the iso-frequency contour. If it is reciprocal, then $ \mathrm{M}_{zz} $ is an even (symmetric) function of in-plane wavenumbers (symmetric iso-frequency contour in momentum domain), otherwise the iso-frequency contour becomes asymmetric in momentum domain. It can be shown that the slope of dissipative decay rate along a specific in-plane direction, let's say $x$-axis, is

\begin{align}
	& \frac{\partial \Gamma_{12}(R)}{\partial x} = \frac{ 2 \mu_{1,z}^2 \omega_0^2 }{\epsilon_0 \hbar c^2}  \partial_x \mathrm{Im} \mathbb{G}_{zz}(\textbf{r}_1, \textbf{r}_2) \nonumber \\&
	\frac{\partial \Gamma_{12}(R)}{\partial x} = \frac{ 2 \mu_{1,z}^2 \omega_0^2 }{\epsilon_0 \hbar c^2} \mathrm{Im}\left[ \frac{i}{8 \pi^2} \int_{k_x} \int_{k_y} (ik_x\mathrm{M}_{zz}(k_x,k_y) )e^{i(k_{x}(x_1 - x_2) + k_y(y_1 - y_2) + k_z(z_1 - z_2)) } dk_x dk_y  \right]
\end{align}
therefore as $ |\textbf{R}| \rightarrow 0 $ the slope of dissipative decay rate becomes

\begin{align}
	\frac{\partial \Gamma_{12}(|\textbf{R}| \rightarrow 0 )}{\partial x} = \frac{ 2 \mu_{1,z}^2 \omega_0^2 }{\epsilon_0 \hbar c^2} \mathrm{Im}\left[ \frac{i}{8 \pi^2} \int_{k_x} \int_{k_y} (ik_x\mathrm{M}_{zz}(k_x,k_y) ) dk_x dk_y  \right]
\end{align}

When electromagnetic reciprocity is respected it can be shown that $ \partial_x \Gamma_{12}(|\textbf{R}| \rightarrow 0 ) = 0 $ because the integrand becomes an odd function of $ k_x $. This indicates in reciprocal case the spontanous emission $ \Gamma_{11} =  \Gamma_{12}(|\textbf{R}| \rightarrow 0 ) $ is an extermum, or, in another word the emitter sits on the maximum of the system Green fuction imaginary part. However, if the reciprocity is violated, then $ \mathrm{M}_{zz} $ is an symmetric function of both $ k_x/k_y $, therefor $ \partial_x \Gamma_{12}(|\textbf{R}| \rightarrow 0 ) \neq 0 $. This reveals that in nonreciprocal case, the slope of Green's function imaginary part is nonzero at the emitter location itself, therefore, $ \Gamma_{11} =  \Gamma_{12}(|\textbf{R}| \rightarrow 0 ) $ is not  an extermum, meaning the maximum of the Green's function imaginary part is pushed away form the emitter location. As a direct result, it becomes possible that $ \Gamma_{12} $ exceeds $ \Gamma_{11} $ for $ \textbf{R} \neq 0 $ which as discussed in the main text, is an essential condition for an efficient inter-atomic photon transport.

Although above discussion proves $ \Gamma_{12} > \Gamma_{11} $ cannot be achieved in planar structurs unless the electromagnetic reciprocity is broken, it is worth mentioning that in reciprocal systems the condition $ \Gamma_{12} > \Gamma_{11} $ can be marginally achieved for very short range interactions, only if the structure contains very sharp discontinuity at the close proximity of the emitters. The inset in Fig. \ref{Figs1} shows two atoms interacting via the reciprocal plasmonic modes of a defected substrate. The defect is relatively large in the order of the emitter radiation wavelength in vacuum $   \approxeq \lambda/2 $. The normalized dissipative decay rate as a function of emitters spacing has been calculated for different emitter-defect distances. The red dot indicate the position of the first emitter at $ x = 0 $. As is clear, only if the defect is at the close proximity of the emitters $ \Gamma_{12} $ barely exceeds $ \Gamma_{11} $ for very small $x$  values (short range interactions). Comparing Fig. \ref{Figs1} with Fig. \ref{Fig2}(d) clearly reveals that, indeed it is violating the reciprocity that can effectively lead to large dissipative decay rate over long range of emitters spacing.

\begin{figure}[bth!]
	\begin{center}
		\noindent
		\includegraphics[width=.45\columnwidth]{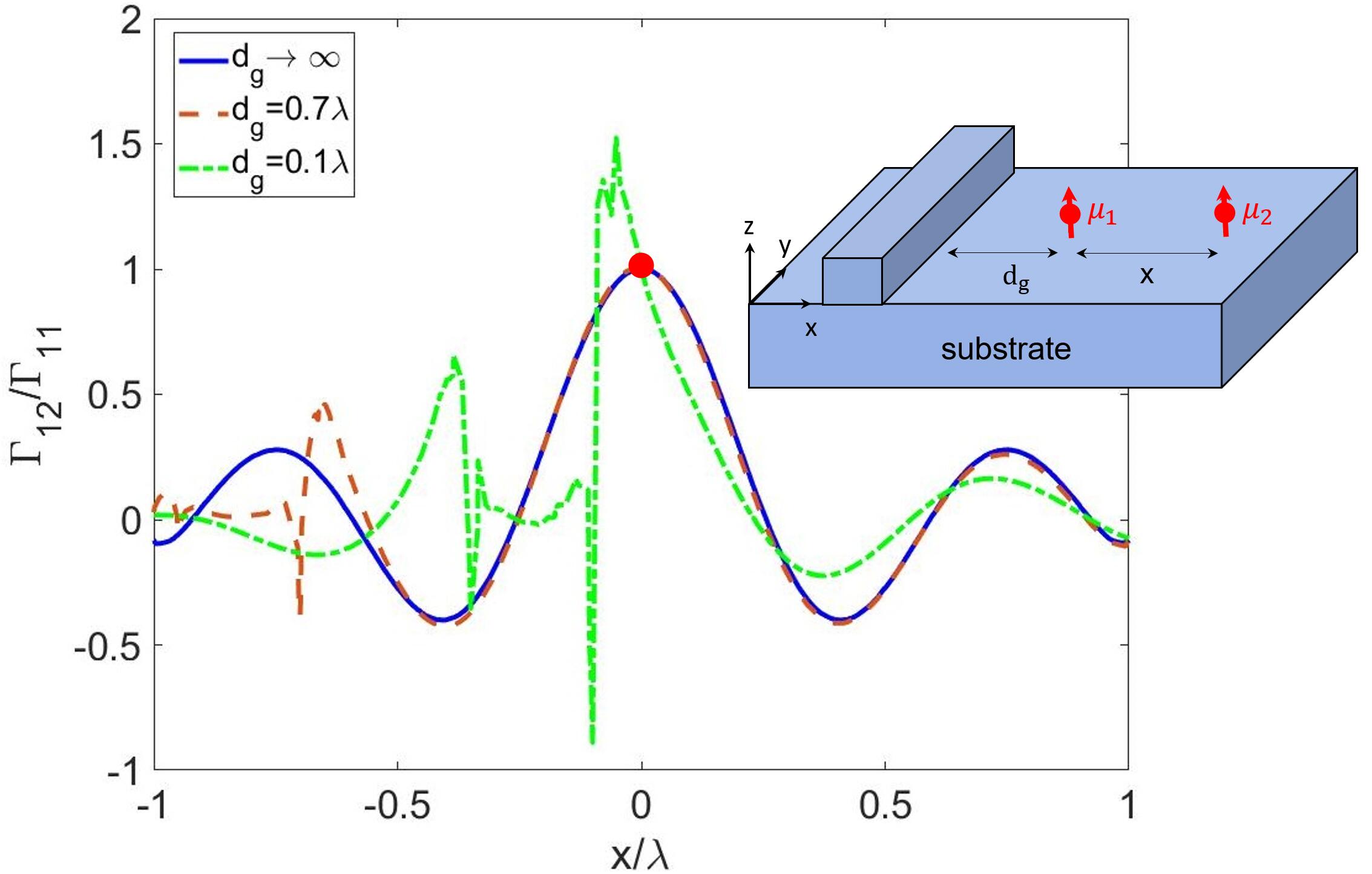}
		\caption{Normalized dissipative decay rate as a function of emitter spacing for different emitter-defect distance. The inset shows the corresponding configuration where two identical emitters interact via reciprcoal plasmonic modes of a defect plasmonic substrate, described by a Drude model permittivity,  $ \epsilon(\omega) = 1 - \left( \omega / \omega_p \right)^2/( 1 + i\gamma/\omega) $. The two emitters (red arrows in the inset) interact at $ \omega / \omega_p = 0.6 $. The defect is relatively large. It is a square with length equal to $ \lambda/2 $, where $ \lambda $ is the radiation wavelength in the free space. The red dot indicate the position of the first emitter at $ x = 0 $.}\label{Figs1}
	\end{center}
\end{figure}

\end{widetext}


\end{document}